\documentclass[aps,prl,twocolumn,showpacs,superscriptaddress]{revtex4}
\usepackage{epsfig}

\def\avg#1{\langle#1\rangle}

\def\be{\begin{equation}} \def\ee{\end{equation}}
\def\bea{\begin{eqnarray}} \def\eea{\end{eqnarray}}

\def\nn{\nonumber}

\begin{document}
\title{Pairing state with a time-reversal symmetry breaking in FeAs-based superconductors}
\author{Wei-Cheng Lee}
\affiliation{Department of Physics, University of California, San Diego,
CA 92093}
\author{Shou-Cheng Zhang}
\affiliation{Department of Physics, McCullough Building,
Stanford University, Stanford CA 94305-4045}
\author{Congjun Wu}
\affiliation{Department of Physics, University of California, San Diego,
CA 92093}

\begin{abstract}
We investigate the competition between the extended $s_{\pm}$-wave
and $d_{x^2-y^2}$-wave pairing order parameters in the iron-based
superconductors. Because of the frustrating pairing interactions
among the electron and the hole fermi pockets, a time-reversal
symmetry breaking $s+id$ pairing state could be favored. We analyze
this pairing state within the Ginzburg-Landau theory, and explore
the experimental consequences. In such a state, spatial
inhomogeneity induces supercurrent near a non-magnetic impurity and
the corners of a square sample. The resonance mode between the
$s_{\pm}$ and $d_{x^2-y^2}$-wave order parameters can be detected
through the $B_{1g}$-Raman spectroscopy.
\end{abstract}
\maketitle

The discovery of the iron-based new superconductors with high
critical temperatures has attracted a great deal of  attention
\cite{kamihara2008,chen_xh2008,chen_gf2008,wen_hh2008,ren2008,rotter2008,
wang2008}. The symmetry structure of Cooper pairs is one of the
central issues for elucidating the superconducting properties. This
problem is complicated by the structure of multiple Fermi surfaces
in LaO$_{1-x}$F$_x$FeAs systems, including hole Fermi surfaces
$\alpha_{1,2}$ around the $\Gamma$-point at $(k_x,k_y)=(0,0)$, and
the electron Fermi surfaces $\beta_{1,2}$ around the M$_{1,2}$
points at $(\pi,0)$ and $(0,\pi)$, respectively. Many theoretical
proposals have suggested the fully-gapped extended $s_{\pm}$-wave
state which preserves the 4-fold rotational symmetry
\cite{mazin2008,kuroki2008,parker2008,chubukov2008,bang2008,seo2008}.
Experimentally, the superfluid density obtained from the penetration
depth measurements is insensitive to temperature, consistent with
this picture \cite{luetkens2008,malone2008,hashimoto2008}. Another
competing pairing structure in the square lattice is $d_{x^2-y^2}$
as proposed by several groups \cite{kuroki2008,seo2008,
chen2008,qi2008}. Kuroki {\it et al.} \cite{kuroki2008} find that
the $d_{x^2-y^2}$ pairing wins over the $s_{\pm}$ state if the
pairing contribution from $\alpha$-Fermi surfaces is suppressed
either by lowering the height of the As atom to the Fe plane
or by heavily electron doping \cite{kuroki2009}.
In particular, several theoretical studies have suggested that $s_{\pm}$ and $d_{x^2-y^2}$ 
pairings are nearly degnerate\cite{kuroki2009,graser2009,zhai2009}.

\begin{figure}
\centering\epsfig{file=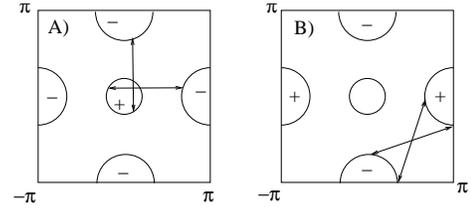,clip=1,width=0.7\linewidth,angle=0}
\caption{The two competing pairing order parameters
LaFeAsO$_{1-x}$F$_x$ systems investigated in Ref.
\cite{kuroki2008,mazin2008} which are compatible with the square
lattice geometry: A) the extended $s$-wave ($s_\pm$) and B) the
d-wave ($d_{x^2-y^2}$). The nesting vectors are indicated by the
bi-directional arrows.
\label{fig:pair}}
\end{figure}

The calculation of the superconducting susceptibility in the five-band model 
shows that the extended $s$-wave and $d_{x^2-y^2}$ pairing orders compete with each
other \cite{kuroki2008}. The spin susceptibility has peaked values
around $(k_x,k_y)=(\pi,0)$ and $(0,\pi)$ which corresponds to the
nesting wavectors connecting $\alpha$ and $\beta$ Fermi surfaces,
and also at nesting wavevectors around $(\pi,\pi/2)$ and $(\pi/2,\pi)$
connecting two $\beta$-surfaces. In the spin fluctuation mechanism,
the pairing order parameters favor opposite signs on two Fermi
surfaces connected by nesting wavevectors. The first nesting favors
the extended $s$-wave pairing, with the opposite signs for the
electron and the hole fermi pockets, and the second one favors the
$d_{x^2-y^2}$-wave pairing, with the opposite signs for the
nearest-neighbor electron pockets, as depicted in Fig.
\ref{fig:pair}. As one can see directly from Fig. \ref{fig:pair},
the pairing interactions based on these two different nesting
vectors lead to a frustrating of the pairing order parameters - the
pure extended $s$ wave and the $d_{x^2-y^2}$ pairing states can not
satisfy both nesting vectors simultaneously. In this situation,
there arises a natural possibility of a mixed $s+id$ pairing state,
which can strike a compromise between the two nesting vectors.

In this article, we investigate the possibility of time-reversal
(TR) symmetry breaking states in LaO$_{1-x}$F$_x$FeAs systems based
on the competition between the extended $s_{\pm}$ and $d_{x^2-y^2}$
order parameters. Based on a Ginzburg-Landau (GL) free energy
analysis, we show that after the occurrence of the $s_\pm$-pairing
at $T_c$, the $s+id$ pairing can develop at a lower temperature
$T^\prime$ by breaking spatial rotation and TR symmetry under
certain conditions. Spatial inhomogeneity can generate supercurrent
around non-magnetic impurities and the corners of square samples due
to a symmetry allowed quadratic gradient coupling of these two order
parameters. The corner tunneling Josephson junction is analyzed. The
resonance mode connecting two pairing order parameters can be
measured through the $B_{1g}$ Raman spectroscopy.

Unconventional superconductivity with TR symmetry breaking effect
has been actively investigated in recent years. Much
experimental evidence has shown that Sr$_2$RuO$_4$ exhibits TR
symmetry breaking effects consistent with the $p_{x}\pm i p_y$
state, including the $\mu$SR and the Kerr effect \cite{luke1998,
xia2006}. Recently, neutron scattering experiments indicate the
existence of the loop current in the pseudogap
region\cite{fauqu2006,varma2006,aji2007} , and the Kerr effect has
also been observed in the YBa$_2$Cu$_3$O$_{6+x}$ system below the
pseudogap temperature \cite{xia2008}. It has been pointed
out that in multi-band superconductors TR symmetry breaking effect
can exist with conventional pairing mechanism due to the repulsive
interband Cooper pair scatterings
\cite{Agterberg1999,tanaka2001,Agterberg2002,ng2008}.
LaO$_{1-x}$F$_x$FeAs systems naturally have multi-band structure,
thus it would be interesting to investigate possible consequences of
the TR symmetry breaking pairing states.

We construct a GL equation to describe the
competition between these two singlet pair channels of the extended
$s$-wave ($s_\pm$) and $d$-wave ($d_{x^2-y^2}$) order parameters as
\bea
F&=&\alpha_s(T) \Delta^*_{s} \Delta_{s}
+ \alpha_d(T) \Delta^*_{d} \Delta_{d}
+ \beta_1 |\Delta_{s}|^4\nn \\
&+&\beta_2 |\Delta_d|^4 +\beta_3 |\Delta_{s}|^2 |\Delta_d|^2
+\beta_4 \Big\{\Delta_{s}^*
\Delta_{s}^*  \Delta_d \Delta_d +c.c.\Big \}
\nn \\
&=&\alpha_s(T) \Delta^*_{s} \Delta_{s}
+ \alpha_d(T) \Delta^*_{d} \Delta_{d}
+ \beta_1 |\Delta_{s^*}|^4+\beta_2 |\Delta_d|^4 \nn \\
&+&(\beta_3 + 2\beta_4) |\Delta_{s^*}|^2 |\Delta_d|^2 -\beta_4 L^2,
\label{eq:GL} \eea where $\alpha_s\approx N_0\ln (T/T_{s})$ and
$\alpha_d\approx N_0 \ln (T/T_d)$; $N_0$ is the density of states at
the Fermi energy; $\beta$s are not sensitive to temperatures;
$L=i(\Delta_s^* \Delta_d -\Delta_d^* \Delta_s)$. Because $\Delta_s$
and $\Delta_d$ belong to different representations of the lattice
symmetry group, they do not couple at the quadratic level but can
couple at the quartic level. In particular, the $\beta_4$-term is
allowed by symmetry because orbital angular momentum is conserved
modulo 4. The thermodynamic stability condition requires that
$\beta_{1,2}>0$ and $4\beta_1\beta_2 -(\beta_3-2|\beta_4|)^2>0$.
$\beta_3$ is also expected to be positive because of the competing
nature between $\Delta_s$ and $\Delta_d$.
A similiar GL theory has been discussed in the high-$T_c$ cuprates\cite{kotliar1988}.

We assume that the pairing tendency in the $s_\pm$ is stronger but
very close to that in the $d_{x^2-y^2}$ channel, i.e., $T_s> T_d$
and $1-T_d/T_s\ll 1$. By minimizing the GL free energy of Eq.
\ref{eq:GL}, the condition for the appearance of $\Delta_d$ at
temperature $T$ is: $-\alpha_d(T) > -\alpha_s(T) \lambda$, 
where $\lambda=(\beta_3-2|\beta_4|)/(2\beta_1)$. This condition can
be satisfied at $\lambda<1$ which gives rise to another critical
temperature $T^\prime$ as: $T^\prime= \sqrt{T_d T_s}
(T_d/T_s)^{\frac{1}{2} \frac{1+\lambda}{1-\lambda}}<T_d$ below
which $\Delta_d$ develops. The sign of $\beta_4$ determines whether
TR symmetry is broken or not. If $\beta_4>0$, it favors a phase
difference of $\pm\frac{\pi}{2}$ between $\Delta_s$ and $\Delta_d$,
i.e., the $s+id$ pairing. On the other hand, the real combination of
$s+d$ is realized at $\beta_4<0$, which preserves TR symmetry. In
both cases, the 4-fold rotational symmetry is broken at $T<T^\prime$
which corresponds to an Ising transition. In the former case a
combined rotation of $90^\circ$ and TR operation still leave the
system invariant.

While the uniform components of the $\Delta_s$ and the $\Delta_d$
components do not couple at the quadratic level, their gradient
terms can. The general gradient terms of the GL free energy are
given by
\bea F_{grad}&=&\gamma_s |\vec \Pi \Delta_s|^2 +\gamma_d
|\vec \Pi \Delta_d|^2
+\gamma_{sd} (\Pi_x^* \Delta_s^* \Pi_x \Delta_d \nn \\
&-& \Pi^*_y \Delta_s^* \Pi_y \Delta_d +c.c.),
\eea
where $\vec \Pi=\vec \nabla - 2 i e \vec A$ and $A$ is the magnetic vector
potential; the $\gamma_{sd}$-term describes the coupling between the
$\Delta_{s,d}$ orders allowed by the square lattice structure
\cite{ren1995}.
Minimizing the free energy we arrive the coupled equations of
\bea
&&\alpha_s \Delta_s + 2\beta_1 |\Delta_s|^2 \Delta_s +\beta_3 |\Delta_d|^2
\Delta_s + 2\beta_4 \Delta_d^2 \Delta_s^* \nn \\
&-&\gamma_s (\Pi_x^2+\Pi_y^2)
\Delta_s -\gamma_{sd}(\Pi_x^2-\Pi_y^2) \Delta_d=0,
\nn \\
&&\alpha_d \Delta_d + 2\beta_2 |\Delta_d|^2 \Delta_d +\beta_3 |\Delta_s|^2
\Delta_d + 2\beta_4 \Delta_s^2 \Delta_d^* \nn \\
&-&\gamma_d (\Pi_x^2+\Pi_y^2)
\Delta_d -\gamma_{sd}(\Pi_x^2-\Pi_y^2) \Delta_s=0.
\label{eq:gl}
\eea
The electric supercurrent can be represented as
\bea
\vec j&=& -\frac{\delta F}{\delta \vec A}=\vec{j}_s+\vec{j}_d+\vec{j}_{ds}
\nn \\
\vec{j}_s&=&2i e \gamma_s [\Delta^*_s \vec \Pi \Delta_s] + c.c. , \nn \\
\vec{j}_d&=&2i e\gamma_d [\Delta^*_d \vec \Pi \Delta_d] +c.c. , \nn \\
\vec{j}_{ds}&=& 2i e\gamma_{sd} \big \{(\Delta^*_s \Pi_x \Delta_d
+ \Delta^*_d \Pi_x \Delta_s) \hat e_x -(\Delta^*_s \Pi_y \Delta_d\nn \\
&+& \Delta^*_d \Pi_y \Delta_s) \hat e_y \big\}+ c.c.,
\label{jds}
\eea
where $\vec j_{s,d}$ are the intra-component supercurrent,
and $j_{sd}$ is the inter-component supercurrent.
The total supercurrent $\vec j$ satisfies the continuity condition: 
$\vec \nabla \cdot \vec j=0$.

\begin{figure}
\centering\epsfig{file=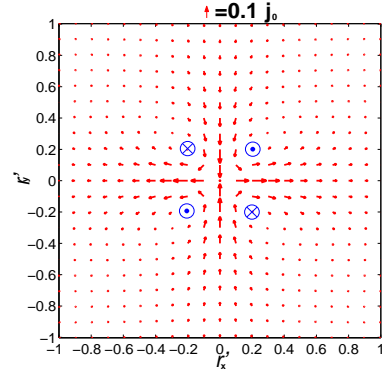,clip=1,width=2.0in,height=2.0in,angle=0}
\caption{\label{fig:current} Supercurrent induced by the impurity at
$\vec{r}=(0,0)$ for $\beta_4>0$. We introduce the length scale
$\xi=\sqrt{\gamma_s/|\alpha_s|}$, energy scale
$\Delta_0=\sqrt{|\alpha_s(T)|/2\beta_1}$, and the density of states
$N_0$. The dimensionless parameters are defined as
$r^\prime_i=r_i/\xi$, $\alpha^\prime_i\equiv \alpha_i/N_0$,
$\beta_i^\prime\equiv \beta_i\Delta^2_0/N_0$,
$\gamma^\prime_i\equiv\gamma_i /\xi^2 N_0$, with the values used
here given by $r_0^\prime=0.1$, $\alpha^\prime_0=10$,
$\alpha^\prime_s(T)=-1.0$, $\alpha^\prime_d(T)=-0.75$,
$\beta^\prime_1=1.0$, $\beta^\prime_2=1.0$, $\beta^\prime_3=0.5$,
$\beta^\prime_4=0.4$, $\gamma^\prime_s=1.0$, $\gamma^\prime_d=0.5$,
and $\gamma^\prime_{sd}=0.35$. The supercurrent is plotted in unit
of $j_0=e\alpha^2_s(T)\xi/\beta_1$ and the length of each arrow is
proportional to the magnitude of the supercurrent. $\odot$ and
$\otimes$ indicate the vorticities of the loop currents appearing in
the four quadrants near the impurity.}
\end{figure}

\begin{figure}
\centering\epsfig{file=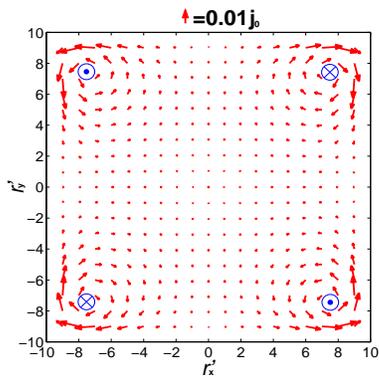,clip=1,width=2.0in,
height=2.0in,angle=0}
\caption{\label{fig:cornercurrent} Supercurrent distribution of a
square sample in terms of the unit of $j_0$ with the same parameters as in Fig.
\ref{fig:current}. Current loops develop with the positive chirality
at the right-down and left-up corners and the negative chirality at
the other two corners.}
\end{figure}

\begin{figure}
\centering\epsfig{file=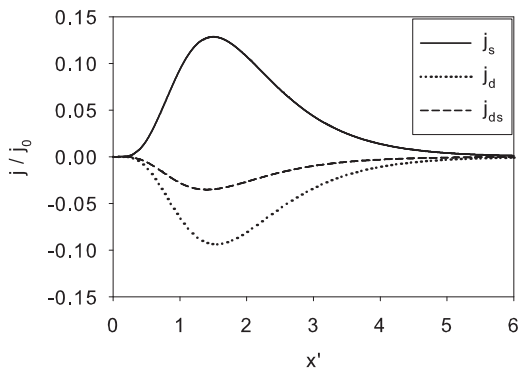,clip=1,width=2.8in,angle=0}
\caption{\label{fig:openedge} Supercurrent components $\vec j_s$,
$\vec j_d$ and $\vec j_{sd}$ in a sample of half plane geometry
($x>0$) plotted with the unit of $j_0$.
The total current $\vec j=\vec j_s +\vec j_d +\vec j_{sd}$ vanishes.
The parameters are the same as in Fig.\ref{fig:current}. }
\end{figure}

A novel consequence of the $\gamma_{sd}$-term is that spatial
inhomogeneity induces supercurrent in a $s+id$ superconductor. We
consider a non-magnetic impurity located at $\vec r_0=(0,0)$ modeled
as a Gaussian distribution of $\alpha$s $\alpha_s(\vec r,T)=
\alpha_{s,d}(T) + \alpha_0 e^{-r^2/r_0^2}$ with $\alpha_0>0$. At
$T<T^\prime$, a $s+id$ superconductor has the relative phase
$\theta_{sd}=\pm\frac{\pi}{2}$ between $\Delta_{s,d}$ in the
homogeneous system. However, in the presence of impurities, the
spatial inhomogeneous distribution of order parameters causes the
inhomogeneity of the relative phase $\theta_{sd}$ due to the
$\gamma_{sd}$ term, which induces supercurrent near the impurity. We
solve Eqs. \ref{eq:gl} and plot the supercurrent pattern in Fig.
\ref{fig:current}. Assuming the magnetic field generated by the
impurity-induced supercurrent is small, we neglect the dependence of
$\vec j$ on the magnetic vector potential $\vec A$. The magnitude of
the supercurrent rapidly decays beyond the order of the healing
length defined $\xi=\sqrt{\frac{\gamma_s}{|\alpha_s|}}$. The
suppercurrent pattern is symmetric under the rotation of
$180^\circ$, and a combined operation of TR and the rotation of $\pm
90^\circ$. Along the $x$ and $y$-axis passing the impurity, the
supercurrents flow along these axis and exhibit the pattern of ``two
in'' and ``two out'', which is consistent with the reflection
symmetry respect to the $x$ and $y$-axis and the continuity
condition. In addition, the system has the combined TR and
reflection symmetry respect to the diagonal axes of $\pm 45^\circ$,
thus supercurrents flow perpendicularly to these axes. This results
in staggered current loops and thus small staggered fluxes in the
four quadrants near the impurity. However, these current loops arise
from both amplitude and phase variations and do not possess
quantized fluxes, and thus are not of topological nature. At
$T_c>T>T^\prime$, a similar effect also occurs since a non-zero
$d$-wave order parameter can be induced despite the supercurrent in
this temperature range is very small.
On the other hand, if a $s+d$ mixing is realized below
$T^\prime$ at $\beta_4<0$, supercurrents will not be induced by
spatial inhomogeneity because $s+d$ does not breaking TR symmetry.
The spatial inhomogeneity only causes the amplitude variation of
$\Delta_{s,d}$, whose relative phase difference remains fixed at
$\theta_{sd}=0$ or $\pi$.

We next consider a square sample and investigate the supercurrent at
the boundary. The spatial variation of $\Delta_{s,d}$ is along
directions normal to edges. The
continuity condition suppresses the supercurrent except at the four
corners as depicted in Fig. \ref{fig:cornercurrent}. Each of four
corners develop a circulating supercurrent loop whose chirality are
staggered as we move around the edges. This is also consistent with
the combined symmetry operation of TR and the rotation of
$90^\circ$. Because at each corner the supercurrent has the same
chirality, thus is easier to be detected by using SQUID than the
single impurity case.

Furthermore, we consider the spatial distribution of $\Delta_{s,d}$ at
$\beta_4>0$ along the edge of a half-plane sample with $x>0$ and the
boundary of the $y$-axis.
Given the boundary conditions of $\Delta_s(x=0,y)=\Delta_d(x=0,y)=0$
and the spatial homogeneity along the $y$-direction, Eq. \ref{eq:gl}
reduces to coupled 1D equations.
Again the continuity condition forbids the appearances of a net supercurrent
for a $s+id$ superconductor, although the three components of the supercurrent,  $j_s$, $j_d$,
and $j_{sd}$, are non-zero individually as depicted in
Fig. \ref{fig:openedge}.
One important feature is that $j_s$ and $j_d$ have {\it opposite} signs.
This internal counterflow supercurrents is a
result of the $\gamma_{sd}$ term.
By plugging $\Delta_{s,d}(x)=\rho_{s,d}(x)e^{i\theta_{s,d}(x)}$ into the GL free energy,
such a term can be approximated as:
$2\gamma_{sd} \sin (\theta_s-\theta_d)[(\partial_x\theta_s)(\partial_x\rho_d)\rho_s-(\partial_x\theta_d)(\partial_x\rho_s)\rho_d]$,
and therefore the free energy can be lowered by choosing
$\mbox{sgn}(\partial_x\theta_s)=-\mbox{sgn}(\partial_x\theta_d)$.

The Josephson junctions touching edges with different orientations of a single
crystal sample act as convincing phase sensitive measurements for the $d$-wave
pairing symmetry in high T$_c$ systems \cite{tsuei2000,
vanharlingen1995} and $p$-wave pairing symmetry in Sr$_2$RuO$_4$
\cite{kidwingira2006,nelson2004}. We consider the same conner
junction in which the two adjacent faces of a single crystal FeAs
superconductor are connected via Josephson weak coupling with a
conventional $s$-wave superconducting thin film. We define that
$J_{s,d}$ are the Josephson coupling between the $s_{\pm} (d)$ order
parameter and the conventional $s$-wave order parameter,
respectively. Assuming that the two junctions at two adjacent faces
are equal, we express the critical current with applied magnetic
flux $\Phi_{ext}$ as\cite{annett1996}:
$I_c(\Phi_{ext})=2I_0\cos\left[\pi\left(\Phi_{ext}/\Phi_0\right)
+\delta/2\right]$, where
$\delta=\tan^{-1}\left[2e_J\sin\theta_{sd}/(1-e_J^2)\right]$,
$e_J\equiv J_d/J_s$, and $\Phi_0$ is the flux quantum. $\delta$
could only be $0$ or $\pi$ for the $s+d$ mixing depending on which
component is dominate, while it can be any value between $0$ and
$\pi$ for the case of $s+id$ mixing.

Now we consider the case of $\langle \hat{\Delta}_s\rangle=\Delta_0$ and $\langle
\hat{\Delta}_d\rangle=0$. In this case we predict that a new
collective mode can be observed in the $B_{1g}$-mode of Raman
spectroscopy, which behaves as the resonance mode connecting
$\hat{\Delta}_{s,d}$, and  is approximately the nematic operator
$\hat{N}_d$ associated with the $B_{1g}$ mode in the Raman
spectroscopy. One can check the following commutator:
$[\hat{N}_d,\hat{\Delta}^\dagger_d] =-2\hat{\Delta}^\dagger_s +
\hat\Delta^{\dagger\prime}_s$, where $\hat{N}_d = 1/V \sum_k
(\cos k_x -\cos k_y) c^\dagger_{\vec k\sigma} c_{\vec k\sigma}$,
$\hat{\Delta}^\dagger_d = 1/V \sum_{\vec{k}}(\cos k_x - \cos
k_y)c^\dagger_{\vec{k}\uparrow} c^\dagger_{-\vec{k}\downarrow}$,
$\hat{\Delta}^\dagger_s = 1/V \sum_{\vec{k}}\cos k_x\cos
k_yc^\dagger_{\vec{k}\uparrow} c^\dagger_{-\vec{k}\downarrow}$, and
$\Delta^{\dagger,\prime}_s= 1/V \sum_{\vec{k}}(\cos^2 k_x+\cos^2 k_y
)c^\dagger_{\vec{k}\uparrow} c^\dagger_{-\vec{k}\downarrow}$. Since
$\avg{\Delta_s}\neq 0$, $\Delta_d$ and $N_d$ are conjugate variables
which lead to collective modes similiar to the $\pi$ resonance mode
in cuprates \cite{demler2004,lee2008}: 
$\omega_{res}\sim  \frac{\sqrt{K^{N_d}K^{\Delta_d}}}{C}$,
where $K^{N_d}\sim N_0\Delta_0^2$ and $K^{\Delta_d}\sim \alpha^*_d(T)
=\alpha_d +(\beta_3-2|\beta_4|)\Delta_0^2$ are the stiffnesses
for $N_d$ and $\Delta_d$ respectively,
and $C\sim N_0$ is the Berry curvature between $N_d$ and $\Delta_d$.
Following the same arguement in Ref. \cite{zhang1997,demler2004},
the excited state with energy $\omega^{res}$ can be defined as:
$\vert \Delta_d\rangle = \hat{\Delta}^\dagger_d\vert 0\rangle$,
where $\vert 0\rangle$ is the BCS ground state.
As a result, at $T=0$ the Raman response function of the $B_{1g}$ mode becomes:
\bea
\chi^{B_{1g}}(\omega)=\sum_n \left\{\frac{\vert\langle 0\vert N_d\vert n\rangle\vert^2}{\omega  - \omega_n + i\epsilon}
-\frac{\vert\langle 0\vert N_d\vert n\rangle\vert^2}{\omega  + \omega_n - i\epsilon}\right\} \nn\\
\approx \left\{\frac{\vert\langle 0\vert [N_d,\hat{\Delta}^\dagger_d]\vert 0\rangle\vert^2}{\omega  - \omega^{res} + i\epsilon}
-\frac{\vert\langle 0\vert [N_d,\hat{\Delta}^\dagger_d]\vert 0\rangle\vert^2}{\omega  + \omega^{res} - i\epsilon}\right\} \nn\\
\sim |\Delta_s|^2 \left\{\frac{1}{\omega  - \omega^{res} +
i\epsilon} -\frac{1}{\omega  + \omega^{res} - i\epsilon}\right\},
\eea which has a sharp peak at $\omega=\omega^{res}$. The temperature
dependence of $\omega^{res}$ may be complex. Nevertheless, the
revelence of $\Delta_d$ can still be infered from it. If the
$\Delta_d$ is competing to $\Delta_s$ but the mixed state is not
favored ($\lambda>1$), $\omega^{res}$ should have weak temperature
dependence and remain finite as $T\to 0$. On the other hand, if the
mixed state could occur at $T=T'$ ($\lambda<1$), $\omega^{res}$
should appear after $T<T_c$ and decrease dramatically to zero as $T$
is approaching $T^\prime$.

In conclusion, we have investigated the competition between the extended
$s$-wave and $d$-wave Cooper pairing orders in the FeAs-based superconductors.
The multiple nesting wavevectors naturally leads to the possibility of
a $s+id$ pairing state which breaks the TR symmetry.
Based on a general Ginzburg-Landau theory we have shown that in such
a state supercurrent can be induced by spatial inhomogeneity, and several
possible experiments to detect this state are discussed.
We also proposed that a new collective mode should be observed in
the $B_{1g}$ Raman spectroscopy as the resonant mode between
the two competing order parameters.

We thank X. Dai, Z. Fang, J. Hirsch, and J. Hu for helpful discussions.
C. W. thanks the Aspen institute of Physics where part of the work was
done.
S. C. Z. is supported by the NSF DMR-0342832 and the US DOE
under Contract No. DE-AC03-76SF00515.
C. W. is supported by the startup fund and the Academic Senate research
grant at UCSD,  the Sloan Research Foundation, and ARO-W911NF0810291.

\end{document}